# THE SMC EXPERIMENT: NEW DATA ON THE DEUTERON FROM THE 1994 RUN

GÜNTER BAUM

*Fakultät für Physik, Universität Bielefeld, D-33501 Bielefeld, Germany*

On behalf of the Spin Muon Collaboration

ABSTRACT

An overview of the SMC data taking and the polarized deep inelastic scattering experiment is given. The new data on the deuteron extend the kinematic range and have considerably reduced statistical and systematic errors. The evaluation of the first moment of the spin dependent structure function is presented and the result for the Bjorken sum rule from SMC data alone is given. The spin contribution of the quarks to the spin of the nucleon is obtained with information from weak decays of baryons. In a new polarized semi-inclusive analysis the asymmetry of the difference between the number of positive and negative charged hadrons was studied. Preliminary results are shown.

## 1. Asymmetries and Spin-Dependent Structure Functions

We consider deep inelastic scattering (DIS) of charged leptons with energy $E$ from a nucleon to energy $E^{'}$ by one-photon exchange. For longitudinally polarized particles the inclusive cross section $\sigma$ is doubly differential and can be described by two kinematic variables, for instance by the square of the momentum transfer, $Q^2 = 4EE^{'} \sin^2 \theta/2$, and the Bjorken scaling variable, $x = Q^2/2M\nu$, where $\nu = (E - E')$ is the energy of the virtual photon. The size of the spin-averaged cross section is determined through the unpolarized structure functions $F_1$ and $F_2$ , which are related by $F_2 = 2x(1 + R)F_1$, with $R$ being a ratio of photo-absorption cross sections, $R = \sigma_L/\sigma_T$. In polarized deep inelastic scattering there appear additional parts of the cross section which modify the scattering and which contain the spin-dependent structure functions, $g_1$ and $g_2$. Due to experimental facts, these modifications typically show up as relative small effects in observations and are best determined by measurements of asymmetries.

To isolate $g_1^d$ for the the deuteron the quantity to be studied experimentally is the asymmetry $A^d$:

$$A^d = \frac{\sigma^{\uparrow\downarrow} - \sigma^{\uparrow\uparrow}}{\sigma^{\uparrow\downarrow} + \sigma^{\uparrow\uparrow}} \tag{1}$$

for antiparallel ($\uparrow\downarrow$) and parallel ($\uparrow\uparrow$) longitudinal spins of the muon and deuteron. A relation can be made to the virtual-photon deuteron asymmetries $A_1^d$ and $A_2^d$:

$$A^d = D(A_1^d + \eta A_2^d) \approx DA_1^d. \tag{2}$$

Extraction of the spin structure function from the expression of the cross section with the definition of Eq. (1) gives

$$g_1^d = \frac{F_2^d}{2x(1+R)} \left( A_1^d + \gamma A_2^d \right) \approx \frac{F_2^d}{2x(1+R)} \frac{A^d}{D}. \quad (3)$$

The coefficients $\eta$ and $\gamma$ depend only on kinematic variables, the depolarization factor for the virtual photon, $D$, in addition on $R$. From the scattering energy and from the size of $A_2^d$, the approximations made in Eq. (2) and in Eq. (3) are very good ones for the SMC experiment. In detail, the virtual-photon deuteron asymmetry $A_1^d$ is defined as

$$A_1^d = \frac{\sigma_0 - \sigma_2}{\sigma_0 + \sigma_2} \quad (4)$$

where 0 and 2 are the total spin projections in the direction of the virtual photon.

Considering the spin dependence in the absorption of photons by quarks $q_i$ of flavour i [$q_i$ = u, d, s] inside the deuteron, a very direct interpretation of $g_1$ can be obtained within the parton model, analogous to the interpretation of $F_1$ in the unpolarized case:

$$g_1^d = \frac{g_1^p + g_1^n}{2} = \frac{1}{4} \left[ \frac{5}{9} (u^+ - u^-) + \frac{5}{9} (d^+ - d^-) + \frac{2}{9} (s^+ - s^-) \right]. \quad (5)$$

$q_i^+$ ( $q_i^-$ ) is the quark distribution with spin parallel (antiparallel) to the spin of the parent deuteron, respectively. We see that

$$\Delta q_i = \int_0^1 ( q_i^+(x) - q_i^-(x) ) \, dx \quad (6)$$

describes the fraction of the nucleon spin carried by the quark flavour $q_i$. Given the above interpretation, it is clear that the first moment of $g_1(x)$, which is the Ellis-Jaffe sum rule for the deuteron, obtains a special significance.

$$\Gamma_1^d = \int_0^1 g_1^d(x) \, dx = \frac{5}{36} \Delta u + \frac{5}{36} \Delta d + \frac{2}{36} \Delta s \quad (7)$$

Here, like in Eq. (5), corrections for the $D$ -state of the deuteron and for QCD effects have been ignored at the moment for simplicity. With information from weak decays of neutrons and hyperons one can extract indirectly values for the individual $\Delta q_i$'s and also for their sum:

$$\Delta\Sigma = \Delta u + \Delta d + \Delta s. \quad (8)$$

$\Delta\Sigma$ describes the fraction of the nucleon spin carried by its quarks. This interpretation and also that of Eq. (6) could however be problematic, due to the axial anomaly in the photon-gluon coupling, if gluons are polarized inside the nucleon.

## 2. Overview of the SMC Data Taking

Measurement of the spin-dependent structure function of the deuteron, $g_1^d$, was first made by the SMC at CERN in 1992 [1]. The target used was basically the EMC setup from the 1984/85 measurements [2], which had, however, been upgraded. This upgrade included installation of a dipol coil, whose transverse field allowed a reversal of the polarization directions by field rotations, thereby minimizing systematic errors. The typical deuteron polarization was about 0.25 until it was discovered that a substantial increase in the polarization could be obtained by rapid frequency modulation. In this way, deuteron polarizations larger than 0.40 were routinely obtained, which provided adequate conditions for successful data taking. The beam energy was 100 GeV (see Table 1 for additional information). The first moment of $g_1^d$ was found to be two standard deviations below the Ellis-Jaffe prediction; the Bjorken sum rule was verified with a precision of 30%, using the EMC proton data [2]. The 1992 deuteron data are in agreement, within their larger errors, with the new 1994 data.

For 1993 polarized protons were chosen as target. The energy of the muon beam was increased to 190 GeV. The SMC experiment is the only polarized DIS setup which has the high energy option. The importance of covering the kinematic region at low-x, that is going to higher energy, was realized. Unfortunately, the doubling of the energy reduces the cross section and thereby the event rate by about a factor two, leading to larger statistical errors.

The results for the longitudinal-spin asymmetry, $A_1^p$, of the SMC [3] are shown in Fig. 1, together with the EMC data [2] and the previous SLAC data [4]. Clearly, the new measurements confirm the earlier experiments, extend their low-x range, and reduce the systematic error size. In particular, the conclusion of the EMC concerning the spin content of the proton is upheld. The structure function $g_1^p$ , extract from $A_1^p$ ( see Eq. (3) ), is also shown in Fig. 1. The evaluation of the first moment at $Q_0^2 = 10$ GeV$^2$ gave $\Gamma_1^p = 0.136 \pm 0.011\,(\text{stat.}) \pm 0.011\,(\text{syst.})$ which is about two standard deviations below the Ellis-Jaffe prediction of $0.171 \pm 0.006$. Using all data available at that time on p, d, and n confirmed the Bjorken sum rule to within 10% of the theoretical value with a precision of 15%.

Also in 1993, the SMC conducted measurements with the target polarized transversely and extracted the asymmetry $A_2^p(x)$. One aim of these exploratory studies was to obtain an experimental upper limit on $A_2^p$, which then also served the purpose to reduce the systematic error on $\Gamma_1^p$. Below $x \leq 0.1$ the SMC values [5] of $A_2^p$ are considerably smaller than the positivity limit of $\sqrt{R}$, and in fact consistent with zero to within $\approx \pm 0.08$. These studies were performed at an energy of 100 GeV.

Table 1. Overview of data taking by SMC in past and future

| Year | $\mu^+$ Energy ( GeV) | Target Material | Target Setup Length $L$ (cm) | $x$ Range | Stat. Error of $\Gamma_1$ ($a^{-1}$) |
|---|---|---|---|---|---|
| 1992 | 100 | d (Butanol) | EMC (old) $L$= 80 | 0.006–0.600 | ±0.02 |
| 1993 | 190 | p (Butanol) | SMC (new) $L$=120 | 0.003–0.700 | ±0.01 |
| 1994 | 190 | d (Butanol) | SMC $L$=130 | 0.003–0.700 | ±0.01 |
| 1995 | 190 | d (Butanol) | SMC $L$=130 | 0.003–0.700 | ±0.01 |
| 1996 | 190 | p ($NH_3$) | SMC $L$=130 | 0.003–0.700 | ±0.008 |

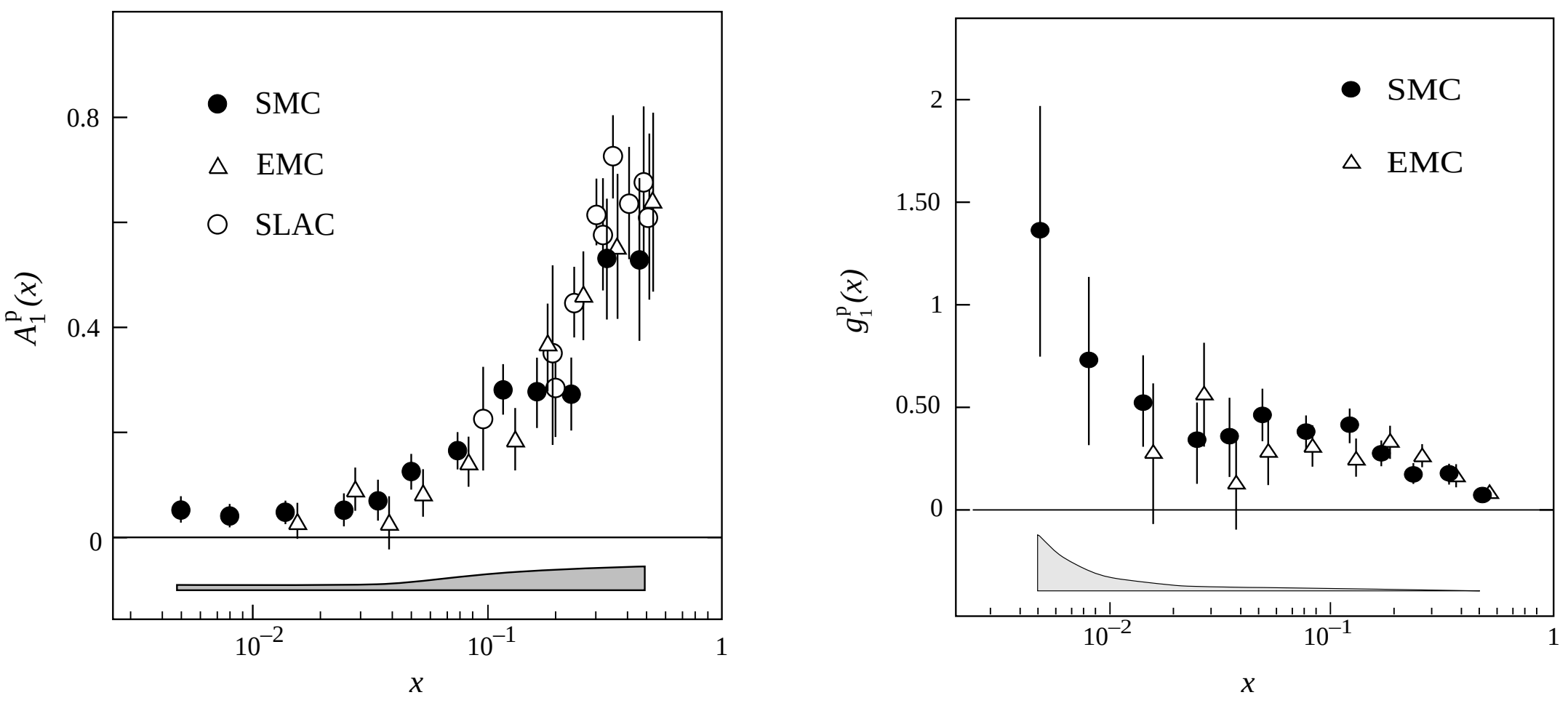

Figure 1. At left the virtual-photon proton cross section asymmetry $A_1^p$ as a function of the Bjorken scaling variable $x$. At right the spin dependent structure function $g_1^p(x)$ at the average $Q^2$ of each x bin; the EMC points are reevaluated using the NMC $F_2$ parametrization [7]. Only statistical errors are shown with the data points. The size of the systematic errors for the SMC points is indicated by the shaded area.

The data obtained for $g_1^d$ in 1994 will be presented in Section 4. Some preliminary results from a new semi-inclusive analysis of SMC polarized DIS, covering 1993 and 1992/94 data using the difference of events of positive and negative charged hadrons will be shown in Section 6. The SMC is taking data on a deuteron target presently (1995). The running will be extended into 1996 with the aim to improve the measurements for the proton [6], with particular focus on the kinematic region at small $x$.

## 3. Experiment

Presently, several experiments are carried out or prepared at CERN with muons, and at SLAC and DESY with electrons to measure the spin structure functions of the nucleons. The advantage and competitive strength of the SMC experiment is its high-energy beam, which allows a unique coverage of the kinematic range of both small $x$ and large $Q^2$ (Fig. 2). As will be seen from the new results (Section 4) this region is of great interest for the behavior of the structure functions themselves and of importance for an accurate determination of the Ellis-Jaffe and Bjorken sum rules. The electron experiments have energies which are less by about a factor four, and hence their minimum values, $x_{min}$, where $x_{min}$ is given by $x_{min} \approx 1/E(\mathrm{GeV})$ are at least four times higher.

The muon beam has, however, a low intensity. Although a thick target can be placed in the beam to increase the luminosity the muon experiment is not competitive with electron scattering experiments in statistical accuracy. Long running times are required to match the systematic error which is of very similar size for all of these spin experiments. Seen altogether, muon and electron experiments can be regarded as complementing each other.

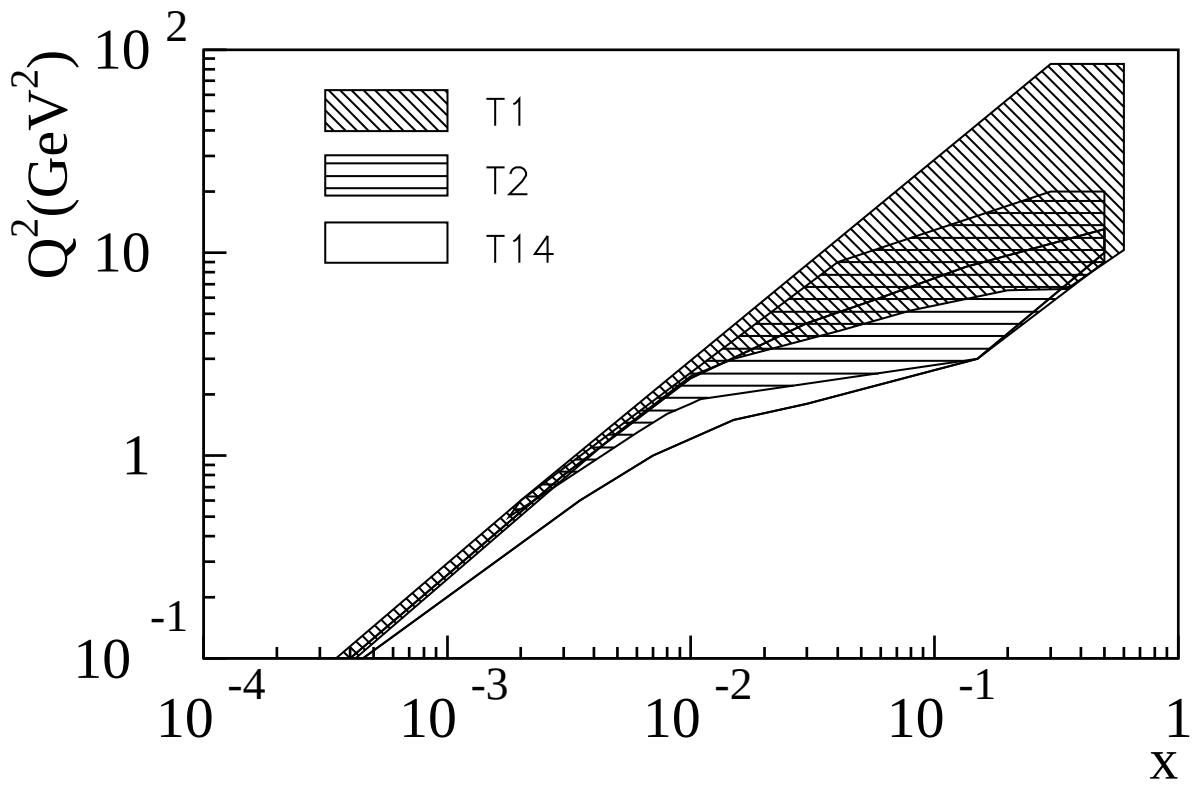

Figure 2. Acceptance for triggers T1, T2, and T14 at 190 GeV/c.

A view of the spectrometer is shown in Fig. 3. For muon tracking it is equipped with over 150 planes of proportional chambers. This is a redundancy which makes the acceptance of the spectrometer largely insensitive to variations of plane efficiencies. This stability is necessary for having the sensitivity to record the small polarization dependent relative scattering differences. Although the spectrometer has been substantially modified since its origin at EMC times, the basic method for obtaining DIS events has not changed [9]. The apparatus contains a calorimeter detector (H2) with electromagnetic and hadron shower part which allows the observation of polarized semi-inclusive scattering events.

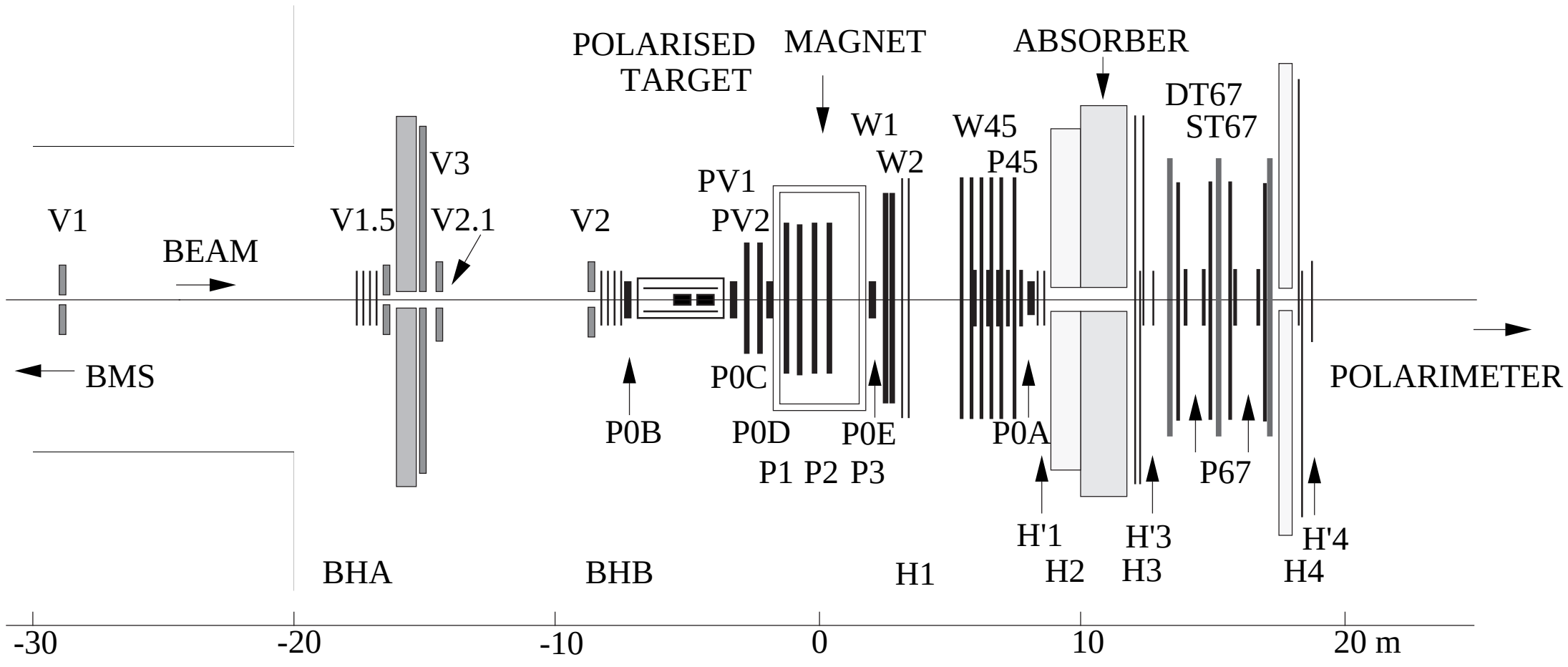

Figure 3. A schematic plane view of the muon spectrometer. BMS: beam momentum spectrometer; V: veto counters for muon halo; BH: muon beam hodoscopes; P: multi-wire proportional chambers; W: drift chambers; H: hodoscope planes for trigger; H2: calorimeter; ST: streamer tubes.

### 3.1. Polarized Muon Beam

The muon beam is polarized naturally due to the properties of weak decays where left handed neutrinos are emitted and spin is conserved in the process $\pi^+ \rightarrow \mu^+\nu$ . Selection of muons in the laboratory frame of reference from the high energy end of the decay spectrum leads to high beam polarization. The polarization was measured with a special polarimeter [10] using the decay of muons to positrons and neutrinos, and was found to be $-0.81 \pm 0.04$, in agreement with Monte Carlo simulation [11]. It should be noted that it is rather impractical to reverse the beam polarization for the asymmetry measurements. A reversal would require a change of the beam line to transport a $\mu^-$ beam, where variations are expected in intensity, location, and phase space of the beam, introducing considerable systematic errors.

### 3.2. Polarized Target

The target is polarized by using the method of dynamic nuclear polarization, involving low temperatures (0.5 K), high magnetic field (2.5 T), microwave irradiation of 70 GHz, and extended NMR measurements [12]. A schematic drawing is shown in Fig. 4 [13]. The beam passes through the inner bore, which is kept as much as possible free of material other than that of the target proper. There are two target cells, an upstream and a downstream one, each 65 cm in length and 5 cm in diameter. The separation between them is 20 cm and they are polarized longitudinally in opposite directions. Together with the vertex resolution capability of $\pm\ 3cm$, this allows spin asymmetry measurements with small systematic errors (see below). Basically, this twin target setup is equivalent to an infinitely fast reversal of a single-cell target. Butanol, which is used as target material [14] has 10 polarizable deuterons (protons) per molecule. For the deuterons a polarization of $P = 0.50$ is obtained now, as average over an expected 150 day long data taking period, for the protons the same quantity is $P = 0.86$.

### 3.3. Method of Data Taking

As the observable asymmetries are quite small it is important to reverse the spin directions of the target halves frequently in order to reduce systematic errors as much as possible. The main method used for this purpose is a rotation of the solenoid field direction by $180^o$ . This is achieved with the help of a superconducting saddle-type dipole coil, wound onto the solenoid [15]. It provides a transverse magnetic field of $0.5\ T$ during the rotation procedure. A rotation was performed every 5 hours during data taking, requiring 10 minutes of interruption when the field was not longitudinally. To guard against possible systematic effects depending on the solenoid field direction, reversals were also performed every two weeks, changing sign of the polarization by interchange of the microwave frequencies for the the target halves. The DIS event

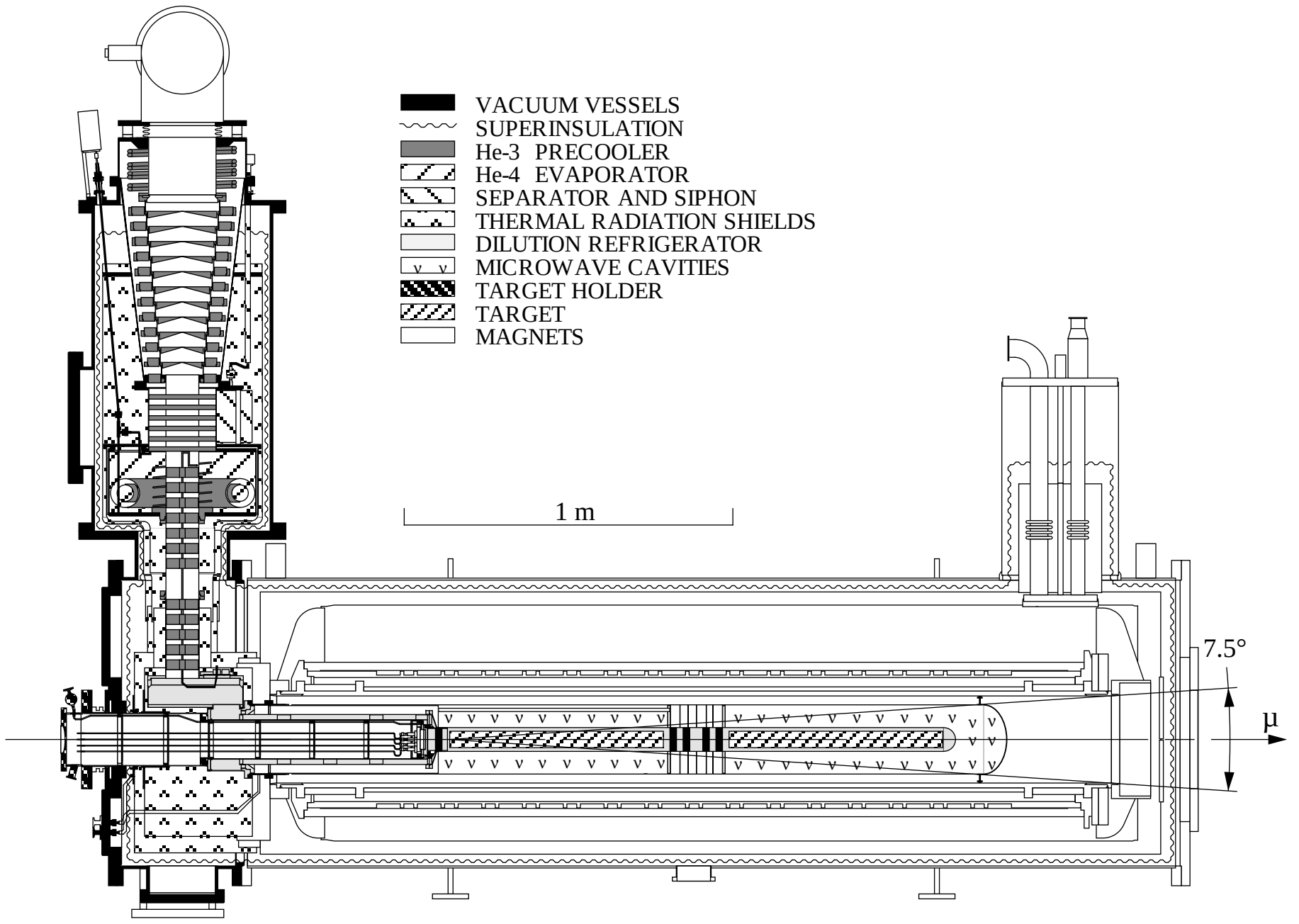

Figure 4. Schematic drawing of the SMC polarized target

yields from the two target halves is given by the following expression:

$$N_{u(d)} = n_{u(d)}\ a_{u(d)}\ \Phi\ \sigma_0\ [\ 1 - f P_\mu P_T D A_1] \tag{9}$$

where the index u refers to the upstream and d to the downstream target cell, n is the number of target nucleons per unit area, the acceptance is denoted by a, $\Phi$ stands for the beam flux, $\sigma_0$ is the unpolarized cross section, f represents the fraction of events coming from polarized nucleons, and $P_T$ is the target polarization. By taking the ratio of the ratio $N_u/N_d$ of events accumulated before a rotation to the same quantity $N_u'/N_d'$ after a rotation,

$$\frac{N_u {N_d}'}{N_d {N_u}'} = (\ 1 +\ 4\ f\ P_\mu\ P_T\ D\ A_1\ ), \tag{10}$$

one sees that dependences on beam flux, target density, and spectrometer acceptance cancel. The latter two only if they are constant in time. The possibility of variations exists for the acceptance ratio, $r = a_u/a_d$, however, the stability can be controlled to a level of $10^{-3}$ leading to false asymmetries $\Delta A_1 \leq 0.01$. The average size of the expression ( $4\ f\ P_\mu\ P_p\ D\ A_1$ ) for the measurements on the proton is about 0.02. For deuterons the expression is smaller because of the lower asymmetry $A_1^d$

## 4. New Results

The new results for $A_1^d(x)$ [16] at the average $Q^2$ of each x bin are shown in Fig. 5. The data points include the events taken previously (see Section 2) with a beam of 100 GeV which have an average $Q^2$ smaller by a factor of about two. The average deuteron polarization for the $6 \times 10^6$ reconstructed events taken in 1994 was 0.48. The dominant systematic errors are coming from uncertainties in $A_2^d$, radiative corrections, target and beam polarizations, time variations of the acceptance ratio r.

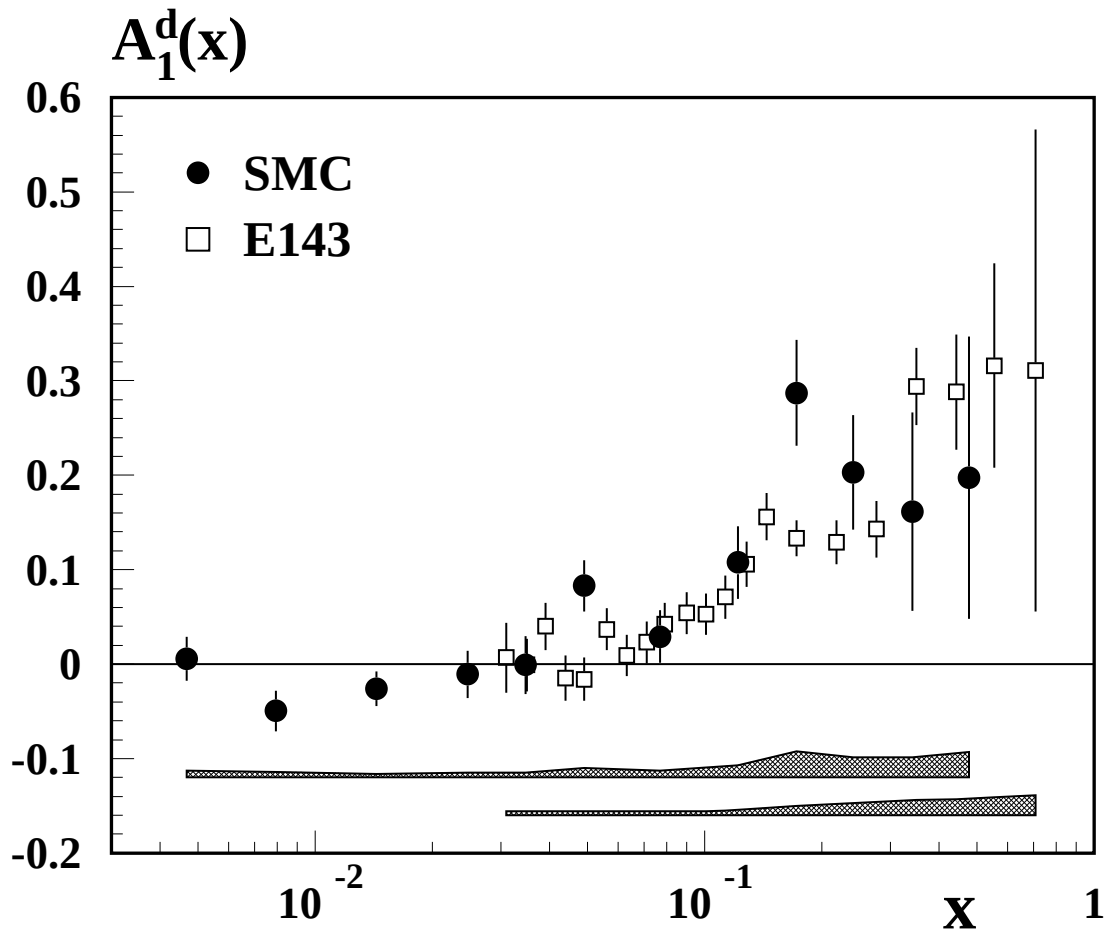

Figure 5. The virtual-photon deuteron cross section asymmetry $A_1^d$ as a function of the scaling variable $x$. Only statistical errors are shown with the data points. The size of the systematic errors is indicated by the shaded area. Results from the SLAC E143 [18] are shown for comparison.

The results for $g_1^d(x)$ obtained from $A_1^d(x)$ are shown in Fig. 6, also at the average $Q^2$ of each x bin, where $Q^2$ varies from 1.3 GeV$^2$ at low x to 48 GeV$^2$ at high x. Below $x \simeq 0.03$ we find $g_1^d$ to be negative.

The spin dependent structure function of the neutron, $g_1^n(x)$, is determined by combining our present result on $g_1^d$ with our $g_1^p$ result [3],

$$g_1^n = \frac{2g_1^d}{1 - \frac{3}{2}\omega_D} - g_1^p, \tag{11}$$

where $\omega_D = 0.05 \pm 0.01$ is the probability of the deuteron to be in a $^3D_1$ -state [17]. We also show $g_1^n$ in Fig. 6. The values below $x \simeq 0.03$ are, like $g_1^d$, negative, more than expected from Regge-type extrapolations from the region above x = 0.03, which is the region in common with the electron scattering experiments at SLAC. E143 obtained results [18] in a similar way like we, E142 obtained results [19] with a $^3$He target. All measurements agree for $g_1$, evolved to a common $Q_0^2$ and within the common x-range. The SMC data show the unexpected effects at small x of a negative $g_1^d$. Together with the unexpected tendency of $g_1^p$ to increase for $x \to 0$, this also results (see Eq. (11) in more negative values for $g_1^n$.

## 5. Evaluation of the First Moment of $g_1^d(x)$ and of $\Delta\Sigma$

The SMC evaluated the first moment of $g_1^d$, which is $\int_0^1 g_1^d(x)dx$, at a $Q_0^2$ of 10 GeV$^2$. It was assumed that the virtual-photon nucleon asymmetries $A_1^d$ do not depend on $Q^2$. According to Eq. (3), $F_2$ and $R$ were taken at $Q_0^2$ from the parametrization of NMC [7] and SLAC [8], respectively. Thus, for each x-bin an evolved $g_1^d$ value was obtained and the integral evaluated. The result obtained is:

$$\Gamma_1^d = 0.034 \pm 0.009 \pm 0.006 \qquad (Q_0^2 = 10 \text{ GeV}^2) \tag{12}$$

where the first error is statistical and the second one systematic. Table 2 states the different contributions to the systematic error. The minor contributions listed as coming from other sources are in order: momentum calibration, uncertainty on $R$, uncertainty on $F_2$, kinematic resolution, extrapolation at high x, dilution factor, proton background. The contribution from the unmeasured region was obtained by extrapolation and amounts to: $-0.002 \pm 0.003$.

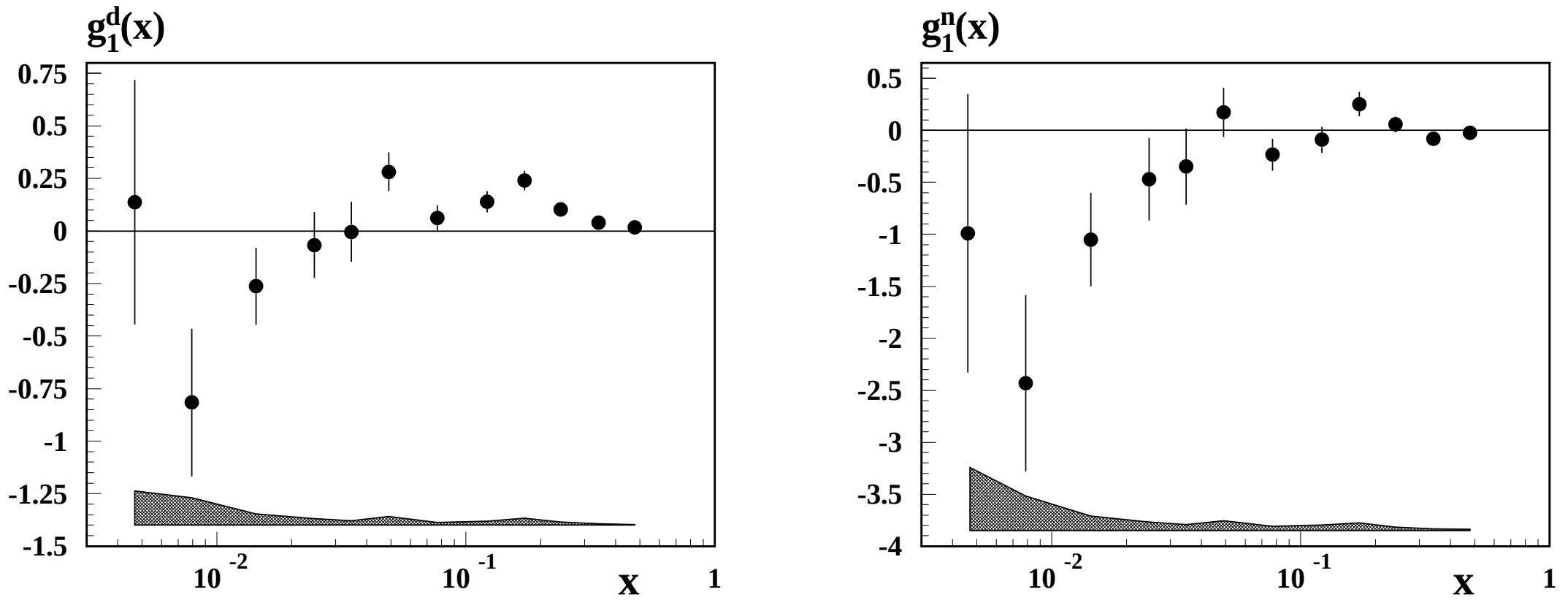

Figure 6. The spin dependent structure functions $g_1^d(x)$ and $g_1^n(x)$, as a function of the scaling variable x at the average $Q^2$ of each $x$ bin as determined by the SMC. Only statistical errors are shown on the data points. The size of the systematic errors is indicated by the shaded area.

The value of $\Gamma_1^d$, corrected for the $D$-state probability with the factor $1/(1 - 1.5\,\omega_D)$, can be used to determine the contributions to the nucleon spin from the sum of the quark spins ($\Delta\Sigma$) and from the strange quark spin ($\Delta s$) alone. Hereby QCD corrections [20] were taken into account [ $C^S(\frac{\alpha}{\pi})$ for the singlet term and $C^{NS}(\frac{\alpha}{\pi})$ for the non-singlet contribution. ]

$$\frac{\Gamma_1^d}{1 - 1.5\,\omega_D} = C^S(\frac{\alpha}{\pi})\frac{1}{9}(\Delta u + \Delta d + \Delta s) + C^{NS}(\frac{\alpha}{\pi})\frac{1}{36}(\Delta u + \Delta d - 2\Delta s) \tag{13}$$

The SMC obtains from their deuteron data alone:

$$\Delta\Sigma = \Delta u + \Delta d + \Delta s = 0.21 \pm 0.11 \tag{14}$$

and

$$\Delta s = -0.12 \pm 0.04. \tag{15}$$

Hereby, the coupling constants for the weak decays of the baryons were taken as $(3F - D) = g_8 = \Delta u + \Delta d - 2\Delta s = 0.579 \pm 0.025$. The QCD corrections were evaluated with $\alpha_s = 0.24 \pm 0.03$ at $Q_0^2 = 10$ GeV$^2$. The results for $\Gamma_1^d$, $\Delta\Sigma$, and $\Delta s$ are inconsistent with the Ellis-Jaffe assumption $\Delta s = 0$ and with the prediction $\Gamma_1^d = 0.070 \pm 0.004$, which is three standard deviations above the measured value. The recent E143 result [18] shows a similar deviation. To test the Bjorken sum rule from the SMC values alone we combine our present result on $\Gamma_1^d$ (Eq. (12)) with our $\Gamma_1^p$ result [3] to obtain

$$\Gamma_1^p - \Gamma_1^n = 0.199 \pm 0.038 \qquad Q_0^2 = 10 \text{ GeV}^2 \tag{16}$$

Table 2. Contributions to the error on $\Gamma_1^d$

| Source of the error | $\Delta\Gamma_1^d$ |
|---|---|
| Extrapolation at low $x$ | 0.0028 |
| Neglect of $A_2$ | 0.0025 |
| Radiative corrections | 0.0025 |
| Beam polarization | 0.0021 |
| Acceptance variation $\Delta r$ | 0.0020 |
| Target polarization | 0.0019 |
| Other (see text) | 0.0029 |
| Total systematic error | 0.0064 |
| Statistics | 0.0088 |

The Bjorken sum rule prediction at $Q_0^2 = 10$ GeV$^2$, including perturbative QCD corrections, is $0.187 \pm 0.003$, in agreement with our value, which has a precision of 19 %.

## 6. Semi-Inclusive Asymmetries

Semi-inclusive measurements in polarized DIS on protons and deuterons can give information on the spin distribution functions of the different flavours for the valence and sea quarks. A determination of four semi-inclusive hadron asymmetries ($A_p^{h^+}, A_p^{h^-}, A_d^{h^+}, A_d^{h^-}$ ) was made by the SMC. These asymmetries were analyzed together with the inclusive asymmetries, $A_1^d$, $A_1^p$ , and the valence, $\Delta u_v(x), \Delta d_v(x)$, and the non-strange sea, $\Delta\overline{q}(x)$, spin distribution functions were extracted. Results have been reported on [21].

Recently, a second method in the analysis of the semi-inclusive data was used with the aim to obtain valence spin distribution functions in a different way, with different systematic errors. Here, the asymmetry for the difference of events yields for positive ($h^+$) and negative ($h^-$) hadrons was studied.

$$A^{h^+-h^-} = \frac{1}{f\ P_\mu\ P_T\ D}\,\frac{(N^{h^+}-N^{h^-})^{\uparrow\downarrow}-(N^{h^+}-N^{h^-})^{\uparrow\uparrow}}{(N^{h^+}-N^{h^-})^{\uparrow\downarrow}+(N^{h^+}-N^{h^-})^{\uparrow\uparrow}}\ , \tag{17}$$

where $f$, $P_\mu$, $P_T$, $D$ are defined in Section 3.3, and $N^h$ are the numbers of events for (anti)parallel orientation of muon and target spin for all hadrons of positive or negative charge.

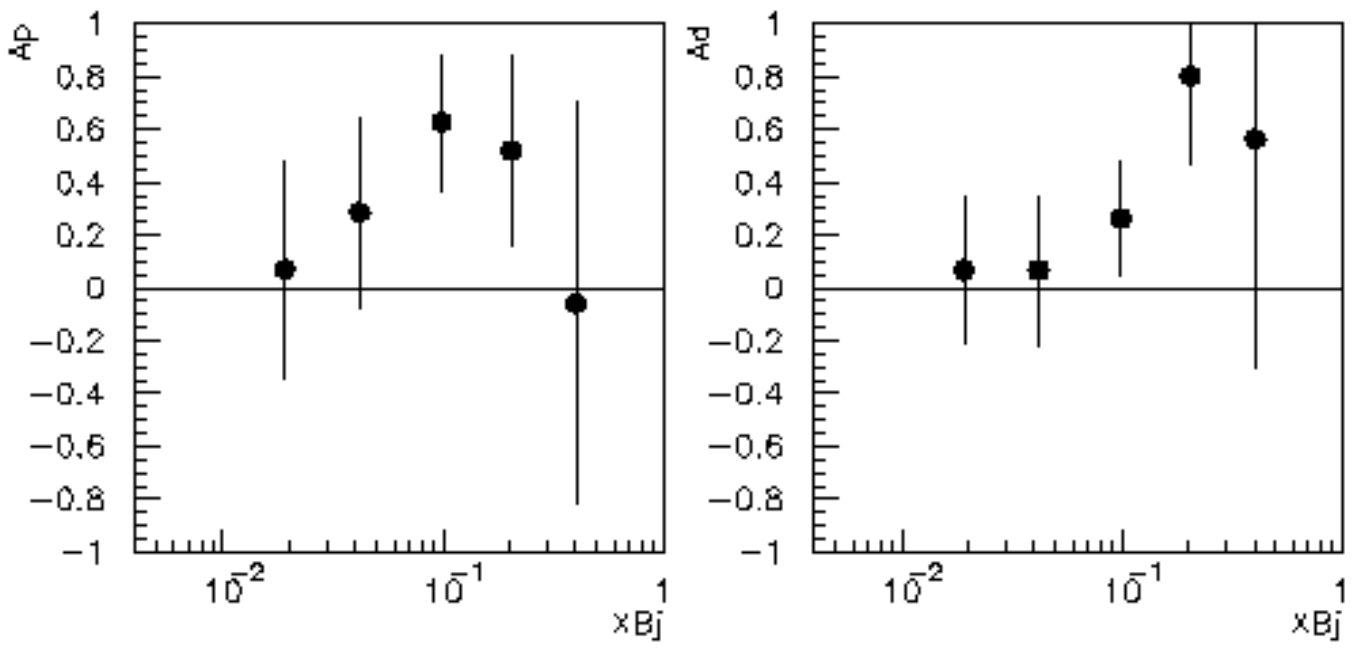

Figure 7. Semi-inclusive asymmetries for the difference of cross sections for positive and negative charged hadrons, $(\sigma^{h^+} - \sigma^{h^-})$. At left: $A_p^{h^+-h^-}$ from the 1993 data; at right: $A_d^{h^+-h^-}$ from the 1992 and 1994 data. Only statistical errors are shown on the data points.

The difference asymmetries do not depend on fragmentation functions and are very simply related to polarized valence quark distributions [22]. For the deuteron

$$A_d^{h^+-h^-} = \frac{\Delta u_v + \Delta d_v}{u_v + d_v} \ , \tag{18}$$

for the proton

$$A_p^{h^+-h^-} = \frac{\Delta u_v - \eta\ \Delta d_v}{u_v - \eta\ d_v}, \tag{19}$$

where $u_v$ ( $d_v$ ) are the unpolarized distribution functions, and the parameter $\eta = 1, 0, -1$ for pions, kaons, protons, respectively.

The experimental signature is that a charged hadron is detected together with a scattered muon. Electrons are rejected through their specific responses in the electromagnetic and hadronic part of the calorimeter H2. The hadrons cannot be identified to be $\pi's, K's$, or $p's$ , however, the sign of the hadron charge is seen by the track curvature in the spectrometer magnet. The measured asymmetries, which have been shown recently [23], are presented in Fig. 7. Cuts were made on the data with $Q^2 \geq 1\mathrm{GeV}^2, x \geq 0.01$, and $z = E_{hadron}/E_{photon} \geq 0.25$.

The analysis does not depend critically on the precise value of $\eta$. For the deuteron ( Eq. (18) ) the asymmetry does not depend on the type of hadron at all. The value of $\eta$ in Eq. (19) is determined from measurements of the EMC, which had a Cherenkov detector for particle identification in the apparatus, and is found to be $\eta = 0.4$ on average. Unpolarized quark distributions are taken at $< Q_0^2 >= 10$ GeV$^2$ from a parametrisation [24].

The polarized quark distributions are shown in Fig. 8. The statistical errors are large, but the results are seen to be significant: $x\Delta u_v(x)$ being relatively large and positive; $x\Delta d_v(x)$ is much smaller than $x\ \Delta u_v(x)$ and seems to be negative at the low-x values. Systematic errors, especially with regard to the first moments of $\Delta u_v(x)$ and $\Delta d_v(x)$ are still being worked on. The addition of future SMC data from the 1995 and 1996 runs will reduce the statistical error size significantly.

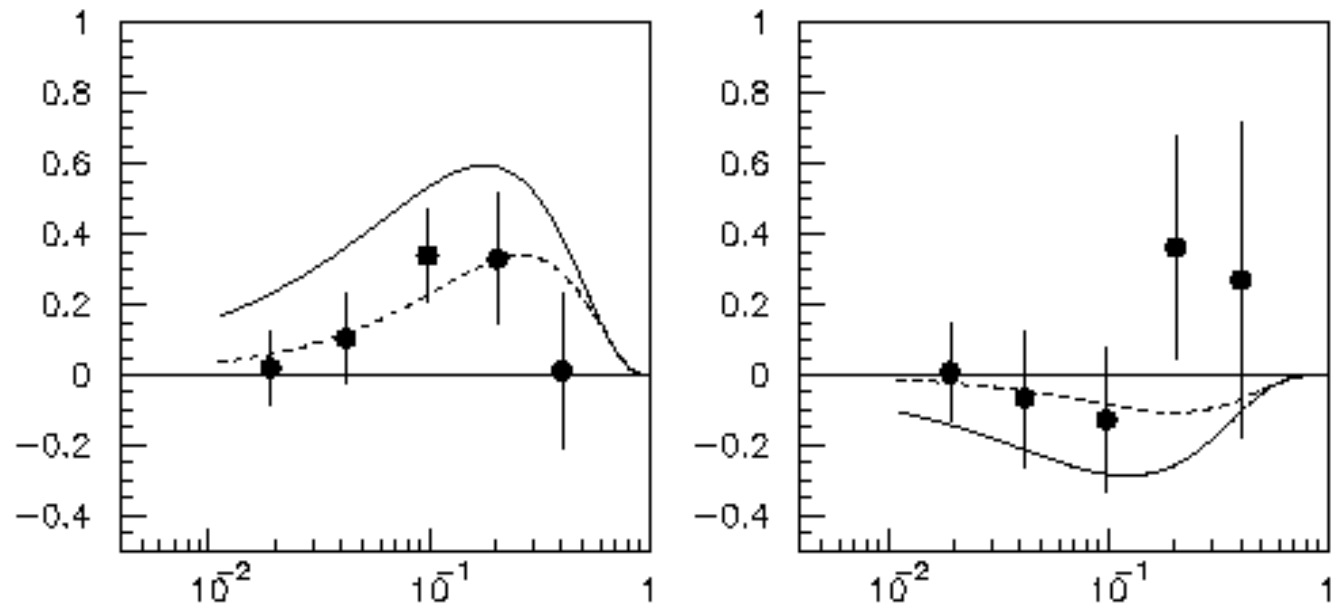

Figure 8. Preliminary results for spin distribution functions $x\Delta u_v(x)$ (left) and $x\Delta d_v(x)$ (right), evaluated from Eq. (18) and Eq. (19). The dashed curve is a parametrization [25]. The full line shows the unpolarized quark distributions at $Q_0^2 = 10\text{GeV}^2$ [24]. Only statistical errors are shown on the data points.

## 7. Summary and Outlook

All the recent results of the polarized DIS experiments on the proton and on the deuteron [3,16,26,18] show a violation of the Ellis-Jaffe prediction for the first moment of the spin structure function for the deuteron as well as for the proton, whereas the Bjorken sum rule is confirmed. These findings lead to similar statements about $\Delta\Sigma$ and $\Delta s$, as the ones that were given above for the SMC results on the deuteron. A first look at semi-inclusive asymmetries has been made. More data about the spin structure of the deuteron and of the proton will come in the near future from further measurements at CERN (SMC) and at SLAC (E154 and E155), as well as at DESY (Hermes). For the more distant future, there are several experiments either under construction (RHIC [27]) or in planning ( HMC [28]) which in particular aim at results for the gluon spin ($\Delta G$) and for the individual quark spin ($\Delta q_i$) contributions. In consideration is also a measurement of the transversity function $h_1(x)$ which gives information about the spin distribution of quarks inside a proton when it is transversely polarized.

## 8. References

1. SMC, B. Adeva et al., *Phys. Lett.* **B302** (1993) 533.
2. EMC, J. Ashman et al., *Phys. Lett.* **B206** (1988) 364; *Nucl. Phys.* **B328** (1989) 1.
3. SMC, D. Adams et al., *Phys. Lett.* **B329** (1994) 399;
   Erratum, *Phys. Lett.* **B339** (1994) 332.
4. SLAC E80, M.J. Alguard et al., *Phys. Rev. Lett.* **37** (1976) 1261; *ibid.* **41** (1978) 70;

SLAC E130, G. Baum et al., *Phys. Rev. Lett.* **51** (1983) 1135.
5. SMC, D. Adams et al., *Phys. Lett.* **B336** (1994) 125.
6. SMC, Addendum to the NA47 Proposal, CERN/SPSCL 95-28.
7. NMC, P. Amaudruz et al., *Phys. Lett.* **B295** (1992) 159 and preprint CERN-PPE/92-124; Errata Oct. 26, (1992) and Apr. 19, (1993).
8. L.W. Whitlow et al., *Phys. Lett.* **B250** (1990) 193.
9. EMC, O.C. Allkofer et al., *Nucl. Instrum. Methods* **179** (1981) 445.
10. SMC, B. Adeva et al., *Nucl. Instrum. Methods* **A343** (1994) 363.
11. N. Doble, L. Gatignon, G. v. Holtey, and F. Novoskoltsev, *Nucl. Instrum. Methods* **A343** (1994) 351.
12. SMC, B. Adeva et al., *Nucl. Instrum. Methods* **A349** (1994) 334.
13. J. Kyynäräinen, *Nucl. Instrum. Methods* **A356** (1995) 47.
14. S. Bültmann et al., *Nucl. Instrum. Methods* **A356** (1995) 102.
15. A. Daël et al., *IEEE Trans. Magn.* **28** (1992) 560
16. SMC, D. Adams et al., CERN-PPE/95-97 .
17. See for instance: R. Machleidt, K. Holinde, and C. Elster, *Phys. Rep.* **149** (1987) 1;
A.Y. Umnikov, F.C. Khanna, and L.P. Kaptari, *Z. Phys.* **A348** (1994) 211 .
18. SLAC E143, K. Abe et al., SLAC-PUB-95-6734 (1995).
19. SLAC E142, D.L. Anthony et al., *Phys. Rev. Lett.* **71** (1993) 959.
20. S.A. Larin, *Phys. Lett.* **B334** (1994) 192.
21. W. Wislicki, in *QCD and High Energy Hadronic Interactions*, ed. J. Trân Thanh Vân, (Editions Frontieres, 1994) p.243 .
22. L.L. Frankfurt et al., *Phys. Lett.* **B230** (1989) 141.
23. F. Perrot-Kunne, Workshop on Deep Inelastic Scattering and QCD, Paris, April 24-28 (1995) .
24. A.D. Martin, W.J. Stirling, and R.G. Roberts, *Phys. Lett.* **B306** (1993) 145.
25. T. Gehrmann and W.J. Stirling, *Z. Phys.* **C 65** (1995) 461.
26. SLAC E143, K. Abe et al., *Phys. Rev. Lett.* **74** (1995) 346.
27. G. Bunce et al., *Particle World* **3** (1992) 1.
28. HMC, Letter of Intent, CERN/SPSLC 95-27.